\documentstyle[preprint,pra,aps,epsf,epsfig,floats,amssymb]{revtex}

\begin{document}


\draft

\title{Interaction-dependent enhancement of the localisation length
  for two interacting particles in a one-dimensional random potential}

\author{Mark Leadbeater,$^1$ Rudolf A. R\"omer,$^2$ and Michael
  Schreiber$^2$}

\address{$^1$Max-Planck-Institut f\"ur Physik komplexer Systeme,
  D-01187 Dresden\\ $^2$Institut f\"{u}r Physik, Technische
  Universit\"{a}t Chemnitz, D-09107 Chemnitz}

\date{$Revision: 2.9 $; compiled \today}

\maketitle

\begin{abstract}
  We present calculations of the localisation length, $\lambda_{2}$,
  for two interacting particles (TIP) in a one-dimensional random
  potential, presenting its dependence on disorder, interaction
  strength $U$ and system size.  $\lambda_{2}(U)$ is computed by a
  decimation method from the decay of the Green function along the
  diagonal of finite samples.  Infinite sample size estimates
  $\xi_{2}(U)$ are obtained by finite-size scaling.  For $U=0$ we
  reproduce approximately the well-known dependence of the
  one-particle localisation length on disorder while for finite $U$,
  we find that $ \xi_{2}(U) \sim \xi_2(0)^{\beta(U)} $ with $\beta(U)$
  varying between $\beta(0)=1$ and $\beta(1) \approx 1.5$. We test the
  validity of various other proposed fit functions and also study the
  problem of TIP in two different random potentials corresponding to
  interacting electron-hole pairs. As a check of our method and data,
  we also reproduce well-known results for the two-dimensional
  Anderson model without interaction.
\end{abstract}

\pacs{71.55.Jv, 72.15.Rn, 71.30.+h}

\maketitle

\narrowtext

%
%

\section{Introduction}

The interplay between disorder and many-body interactions in
electronic systems has been studied intensively over the last two
decades \cite{LR85} and still continues to receive much attention.
Unlike the case of non-interacting electrons, where the ``scaling
hypothesis of localisation'' \cite{AALR79} can reliably predict the
results of many experimental and numerical studies, there is no
equally successful approach to localisation when many-particle
interactions become important \cite{LR85}.  Recently, experimental
studies of persistent currents in mesoscopic rings and the discovery
of a metal-insulator transition in certain two-dimensional (2D)
electron gases at zero magnetic field \cite{KSS96} have shown that the
presence of interactions can indeed give rise to both quantitatively
and qualitatively unexpected phenomena.

A simple and tractable approach to the problem of interacting
electrons in disordered materials is the case of only two interacting
particles (TIP) in a random potential in one dimension.  For a Hubbard
on-site interaction this problem has recently attracted a lot of
attention after Shepelyansky \cite{S94,S96} argued that attractive as
well as repulsive interactions between the two particles (bosons or
fermions) lead to the formation of particle pairs whose localisation
length $\lambda_2$ is much larger than the single-particle (SP)
localisation length $\lambda_1$ \cite{CKM81,KW81}.  Based on a mapping
of the TIP Hamiltonian onto an effective random matrix model (RMM) he
predicted
\begin{equation}
  \lambda_2 \sim (U/t)^2 \lambda_1^2
  \label{eq-shep}
\end{equation}
at two-particle energy $E=0$, with $t$ the nearest-neighbor hopping
matrix element and $U$ the Hubbard interaction strength.  Shortly
afterwards, Imry \cite{I95} used a Thouless-type block-scaling picture
(BSP) in support of this. The major prediction of this work is that in
the limit of weak disorder a particle pair will travel much further
than a SP.  This should be contrasted with renormalization group
studies of the 1D Hubbard model at finite particle density which
indicate that a repulsive onsite interaction leads to a strongly
localised ground state \cite{GS88}.

The preferred numerical method for accurately computing localisation
lengths in disordered quantum systems is the transfer matrix method
(TMM) \cite{MK83}.  Thus it was natural that the first numerical
studies devoted to the TIP problem also used the TMM to investigate
the proposed enhancement of the pair localisation length $\lambda_2$
\cite{S94,FMPW95}.  Other direct numerical approaches to the TIP
problem have been based on the time evolution of wave packets
\cite{S94,BGKTMV98}, exact diagonalization \cite{WMPF95}, or Green
function approaches \cite{OWM96,SK97}.  In these investigations
usually an enhancement of $\lambda_2$ compared to $\lambda_1$ has been
found but the quantitative results tend to differ both from the
analytical prediction in Eq.\ (\ref{eq-shep}), and from each other.
Furthermore, a check of the functional dependence of $\lambda_2$ on
$\lambda_1$ is numerically very expensive since it requires very large
system sizes.

Following the approach of Ref.\ \onlinecite{FMPW95}, two of us studied
the TIP problem by a different TMM \cite{RS97} at large system size
$M$ and found that (i) the enhancement $\lambda_2/\lambda_1$ decreases
with increasing $M$, (ii) the behavior of $\lambda_2$ for $U=0$ is
equal to $\lambda_1$ in the limit $M\rightarrow\infty$ only, and (iii)
for $U\neq 0$ the enhancement $\lambda_2/\lambda_1$ also vanishes
completely in this limit.  Therefore we concluded \cite{RS97} that the
TMM applied to the TIP problem in 1D measures an enhancement of the
localisation length which is due to the finiteness of the systems
considered.
The main problem with this approach is that the enhanced localisation
length $\lambda_2$ is expected to appear along the diagonal sites of
the TIP Hamiltonian, whereas the TMMs of Refs.\ \cite{FMPW95,RS97}
proceed along a SP coordinate.  Various new TMM approaches have
been developed to take this into account
\cite{FMPW95,RS97,RS98,HZKM98}, but still all TMMs share a common
problem: in general the $U=0$ result for $\lambda_2$ does not equal
the value of $\lambda_1/2$ which is expected for non-interacting
particles as explained below.  Rather, they show localisation lengths
$\lambda_2(U=0)$ which are much larger than $\lambda_1/2$ and very
close to $\lambda_2( U \approx 1 )$.

The obvious failure of the TMM approach to the TIP problem in a random
potential has lead us to search for and apply another well-tested
method of computing localisation lengths for disordered system: the
decimation method (DM) \cite{LW80}. Furthermore, instead of simply
considering localization lengths $\lambda_2(U)$ obtained for finite
systems \cite{FMPW95,WMPF95,OWM96,SK97}, or by simple extrapolations
to large $M$ \cite{RS97}, we will construct finite-size scaling (FSS)
curves and compute from these curves scaling parameters which are the
infinite-sample-size estimates of the localization lengths $\xi_2(U)$.
We find that onsite interaction indeed leads to a TIP localisation
length which is {\em larger} than the SP localisation length at $E=0$
and for not too large $U$.  However, the actual functional dependence
is not simply given by Eq.\ (\ref{eq-shep}). In fact our data allow us
to see $\xi_2(U) \sim \xi_2(0)^{\beta}$ with an exponent $\beta$ which
increases with increasing $|U|$ at $E=0$.

The paper is organized as follows: In section \ref{sec-dm} we
introduce the numerical DM used to compute the localisation lengths.
In section \ref{sec-test}, we investigate the numerical reliability of
the DM by studying the Anderson model in 2D.  We then apply the method
to the case of TIP in section \ref{sec-tipdm} and use FSS in order to
construct infinite-sample-size estimates in section
\ref{sec-tipdm-fss}.  We fit our data with various functional forms
for $\xi_2$ put forward in the literature. In section \ref{sec-iehdm}
we also study the problem of two interacting particles in different
random potentials. In section \ref{sec-ab2ddm}, we study the related
problem of a SP in a 2D random potential with additional barriers. We
conclude in section \ref{sec-concl}.

%
%

\section{The Decimation method}
\label{sec-dm}

We shall be considering properties of Hamiltonians of the form
\begin{eqnarray} 
{\bf H} & = &
- t \sum_{n,m} \left( \vert n, m \rangle\langle n+1,m\vert + 
                      \vert n, m \rangle\langle n,m+1\vert + h.c. \right)
\nonumber \\
 & & \mbox{ }
+ \sum_{n, m} \vert n, m \rangle 
              \left( \epsilon^1_n + \epsilon^2_m + U(n)\delta_{nm} \right) 
              \langle n,m\vert
\label{eq-ham}
\end{eqnarray} 
where the choice of $\epsilon^1_n$, $\epsilon^2_m$ and the definition
of $U(n)$ depends on the specific problem considered. For the case of
TIP in 1D the indices $n$ and $m$ correspond to the positions of each
particle on a 1D chain of length $M$ and $\epsilon^1_n =\epsilon^2_n
\in [-W/2,W/2]$. We shall also present results for the case of
$\epsilon^1_n \neq \epsilon^2_n$ which corresponds to two interacting
particles in different 1D random potentials, e.g., two electrons on
neighboring chains, or an electron and a hole on the same chain. In
these cases $U(n)=U$ is the interaction between the two particles.
Instead of considering TIP we can also choose $M^2$ uncorrelated
random numbers $\tilde{\epsilon}_{nm}\in [-W/2,W/2]$ and replace
$\epsilon^1_n+\epsilon^2_m$ in (\ref{eq-ham}) by
$\tilde{\epsilon}_{nm}$. Then the Hamiltonian (\ref{eq-ham})
corresponds to the standard Anderson model for a single particle in 2D
with an additional potential $U(n)$ along the diagonal of the 2D
square.  In all cases we use hard-wall boundary conditions and $t
\equiv 1$ sets the energy scale.

We now proceed to construct an effective Hamiltonian along the
diagonal of the $M\times M$ lattice by using the DM \cite{LW80}. If we
write ${\bf A}(E) = E{\bf 1}-{\bf H}$, the defining equation ${\bf
  A}(E) {\bf G}(E) = {\bf 1}$ for the Green function ${\bf G}(E)$ can
be written as
\begin{equation} 
\sum_{j=1}^{N-1}
  A_{ij}(E)G_{jk}(E) + A_{iN}(E)G_{Nk}(E)  = \delta_{ik}
\label{eq-dec}
\end{equation} 
where $N=M^2$ is the total number of sites in the system and the
indices $i,j,k= 1, \ldots, N$ represent multi-indices for the $M^2$
states $|n,m\rangle$.  From this we can see by choosing $i=N$ that
\begin{equation} 
G_{Nk}(E) =
{\delta_{Nk} \over A_{NN}(E)} 
- \sum_{j=1}^{N-1} {A_{Nj}(E) \over A_{NN}(E)}G_{jk}(E).  
\end{equation} 
Substituting into (\ref{eq-dec}) gives for all $k \neq N$
\begin{equation}
  \sum_{j=1}^{N-1} \left[A_{ij}(E) - {A_{iN}(E)A_{Nj}(E)\over
      A_{NN}(E)}\right]G_{jk}(E)  = \delta_{ik}.
\end{equation}
In this way we have obtained an effective Hamiltonian ${\bf H}'(E)$
with matrix elements $H'_{ij}(E) = H_{ij} + {H_{iN}H_{Nj}\over
  E-H_{NN}}$ whose Green function is identical to that of the full
Hamiltonian on all non-decimated sites.  This process is repeated
until we are left with an effective Hamiltonian for the
doubly-occupied sites only.
We remark that due to cpu-time considerations it turned out to be
useful to split the Hamiltonian into two halves along the diagonal and
to start the decimation process from the outer corner of the
triangular half and then decimate in slices towards the diagonal.  The
procedure is shown pictorially in Fig.\ \ref{fig-dm}. Furthermore, for
the case of TIP we only need to decimate one half and can use the
symmetry of the spatial part of the wave function for the other half.

We shall now focus our attention upon the TIP localisation length
$\lambda_2$ obtained from the decay of the transmission probability of
TIP from one end of the system to the other. In accordance with the SP
case \cite{MK83}, $\lambda_2$ is defined by the TIP Green function,
${\bf G_2}(E)$. More precisely \cite{OWM96}
\begin{equation}
  {1\over\lambda_2} = - {1\over\vert M-1\vert} \ln\vert\langle
  1,1\vert {\bf G_2}\vert M,M\rangle\vert.
\label{eq-lambda2}
\end{equation}
The Green function matrix elements $\langle 1,1\vert {\bf G_2}\vert
M,M\rangle$ are computed by inverting the matrix $\tilde{{\bf A}}(E) =
E{\bf 1}-\tilde{{\bf H}}(E)$ obtained from the effective Hamiltonian
$\tilde{{\bf H}}(E)$ for the doubly occupied sites. We remark that in
order to reduce possible boundary effects, we compute $\lambda_2$ by
considering the decay between sites slightly inside the sample instead
of the boundary sites $(M,M)$.

%
%

\section{Testing the Decimation Method}
\label{sec-test}

As mentioned in the introduction, one of the surprises of the TIP
problem is the apparent inapplicability of the TMM approach, which
leads to large enhancement of the localisation lengths even in the
absence of interaction ($U=0$).  Thus it appears necessary that before
using the DM for the case of TIP, we should also check that by
restricting ourselves to the decay of the Green function along the
diagonal, we do not encounter similar artificial enhancements of
$\lambda_2(U=0)$. As a first test, we have therefore studied the decay
of the Green function along the diagonal for the usual 2D Anderson
model at various disorders $W= 0.65, \ldots, 20$ and system sizes $M=
51, \ldots, 261$. For comparison, estimates of $\lambda_1$ were
computed by the standard TMM \cite{MK83} in 2D.  We then use FSS as in
\cite{MK83} and compute the localisation lengths $\xi_1(W)$ valid at
infinite system size for both sets of data.

In Fig.\ \ref{fig-a2ddm-xi} we show the resulting localisation lengths
$\xi_1$ obtained by TMM with $1\%$ accuracy and DM averaged over 100
samples for each $W$ and $M$.
When we are considering a 2D system, to obtain the correct value of
the localisation length we have to multiply the localisation length
obtained from Eq.\ (\ref{eq-lambda2}) by $\sqrt{2}$ to take account of
the fact that we are studying the decay along the diagonal.
As shown in Fig.\ \ref{fig-a2ddm-xi}, the agreement is good down to
$W=4.5$ where the FSS becomes unreliable. We clearly see that using
the DM to calculate the Green function along the diagonal reproduces
the well-known results of Ref.\ \onlinecite{MK83} up to a geometrical
factor which is easily understood. Furthermore, the deviations from
the TMM results for $\xi_1$ show that our method underestimates the
infinite system size results.  Therefore the above mentioned problem
of the TMM giving rise to too large a value for the TIP-localisation
lengths $\lambda_2$ due to small system size should not appear. We
emphasize that the FSS procedure is more than an extrapolation to the
infinite system size \cite{MK83} and it allows us to identify the
disorders at which FSS breaks down as shown in Fig.\ 
\ref{fig-a2ddm-xi}.

Before proceeding to the case of TIP, we need to discuss an important
difference between the data obtained from TMM and DM. The TMM proceeds
by multiplying transfer matrices for 2D strips (3D bars) of finite
size $M$ ($M\times M$) many times until convergence is achieved. The
localisation lengths are then computed as eigenvalues of the resulting
product matrix \cite{MK83}. However, in the present case of DM (or any
other Green function method applied to TIP), the localisation lengths
are estimated by assuming an exponential decay as in Eq.\ 
(\ref{eq-lambda2}). Such a simple functional form, however, will no
longer be reliably observable when $\xi_1 \sim M$ and we will start to
measure the oscillations in the Green function underlying the
exponential envelope (\ref{eq-lambda2}). Looking at Fig.\ 
\ref{fig-a2ddm-xi}, we indeed see that the deviations from the TMM
result start at $\xi_1 \approx 250$, that is, just at the largest
system sizes used. Increasing the number of samples will reduce this
effect, but this quickly becomes prohibitive due to the immense
computational effort.  With this in mind, we now continue to the case
of TIP.

%
%

\section{The TIP problem at fixed system size}
\label{sec-tipdm}

We now compute the Green function at $E=0$ for 26 disorder values $W$
between $0.5$ and $9$ indicated in Fig.\ \ref{fig-tipdm-l2_w}, for 24
system sizes $M$ between $51$ and $251$, and 11 interactions strengths
$U= 0, 0.1, \ldots, 1.0$. For each such triplet of parameters
$(W,M,U)$ we average the inverse localisation lengths $1/\lambda_2$
computed from the Green function according to Eq.\ (\ref{eq-lambda2})
over 1000 samples.

In Fig.\ \ref{fig-tipdm-l2_w}, we show the results for $M= 201$. Let
us first turn our attention to the case $U=0$. As pointed out
previously \cite{SK97}, the TIP Green function ${\bf G_2}$ at $E=0$ is
given by a convolution of two SP Green functions ${\bf G_1}$ at
energies $E_1$ and $-E_1$ as
\begin{equation}
 \langle 1,1 | {\bf G_2}(0) | M,M \rangle \sim
 \int dE' \langle 1 | {\bf G_1}(E') | M \rangle
          \langle 1 | {\bf G_1}(-E') | M \rangle .
\label{eq-g1g1}
\end{equation}
Assuming that $\langle 1 | {\bf G_1}(E) | M \rangle \propto \exp
\left[ - | M - 1 | / \lambda_1(E) \right]$, where $\lambda_1(E)$ is
the SP localisation length of states in the 1D Anderson model
\cite{KW81}, one expects that the largest localisation lengths
dominate the integral.  Since $\lambda_1(0) \geq \lambda_1(E)$, this
implies that $\langle 1,1 | {\bf G_2}(0) | M,M \rangle \approx \exp
\left[ - 2| M - 1 | / \lambda_1(0) \right]$. Applying Eq.\ 
(\ref{eq-lambda2}), we get $\lambda_2 = \lambda_1/2$ \cite{remark}.
Therefore we have also included data for $\lambda_1/2$ in Fig.\ 
\ref{fig-tipdm-l2_w}.  Since $\lambda_1$ deviates from the simple
power-law prediction \cite{KW81} $\lambda_1 \approx 104/W^2$ at $E=0$
for $\lambda_1 \lesssim 4$ ($W\gtrsim 5$), we have computed
$\lambda_1$ by TMM \cite{CKM81} in 1D with $0.1\%$ accuracy.

Comparing these results with the TIP localisation lengths obtained
from the DM, we find that for $1 \leq W \leq 6$, the agreement between
$\lambda_2(U=0)$ and $\lambda_1/2$ is rather good and, contrary to the
TMM results, there is no large artificial enhancement at $U=0$. For
smaller disorders $W < 1$, we have $\lambda_2 \approx M/2$ so that it
is not surprising that the Green function becomes altered due to the
finiteness of the chains \cite{VRS97}.  This results in reduced values
of $\lambda_2$. For large disorders $W > 6$, we see a slight upward
shift of the computed $\lambda_2$ values compared to $\lambda_1/2$.
This effect is due to a numerical problem, since straightforward
application of Eq.\ (\ref{eq-lambda2}) is numerically unreliable for
values of $\lambda_1$ as small as 1.

It is noticeable from these results, however, that the values of
$\lambda_2(U=0)$ are still slightly larger than $\lambda_1/2$. In
order to explain this behavior, we have computed $\langle 1 | {\bf
  G_1}(E) | M \rangle$ by exact diagonalization of the SP Hamiltonian
for at least 100 samples at many different energies inside the band
and then integrated as in Eq.\ (\ref{eq-g1g1}). Plotting the resulting
localisation lengths in Fig.\ \ref{fig-tipdm-l2_w} we see that indeed
the agreement with $\lambda_2(U=0)$ is better than with $\lambda_1/2$.
Thus the corresponding conjecture of Ref.\ \cite{SK97} is shown to be
true.

For $U$ between $0.1$ and $1$ we have found that the localisation
lengths are increased by the onsite interaction (cp.\ Fig.\ 
\ref{fig-tipdm-l2_w}). We have also seen that for $W > 1.4$ the
localisation lengths $\lambda_2(U)$ increase with increasing $U$. For
smaller $W$ we have $\lambda_2(U) \sim M/2$ and, as discussed above,
the data become unreliable for fixed system size.

Up to now we have been mostly concerned with the behavior of
$\lambda_2$ as function of disorder for $U \in [0,1]$.  However, for
large $U$, it is well-known that the interaction splits the single TIP
band into upper and lower Hubbard bands. Thus we expect that for large
$U$ the enhancement of the TIP localisation length vanishes. In Fig.\ 
\ref{fig-tipdm-l2_u} we present data for $\lambda_2(U)/\lambda_2(0)$
obtained for three different disorders for system sizes $M=201$ at
$E=0$. In agreement with the previous arguments and calculations
\cite{RS97,PS97,WWP98}, we find that the enhancement is symmetric in
$U$ and decreases for large $|U|$. For small $|U|$, we see that the
localisation length increases nearly linearly in $|U|$ with a slope
that is larger for smaller $W$. We do not find any $U^2$ behavior as
in Refs.\ \cite{S94,S96,I95}. In Ref.\ \cite{WWP98} is has been argued
that at least for $\lambda_1 \approx M$, there exists a critical
$U_c={24}^{1/4}$, which is independent of $W$, at which the
enhancement is maximal. We find that in the present case with
$\lambda_1 < M$ the maximum enhancement
$\max_{U}\left[\lambda_2(U)/\lambda_2(0)\right]$ depends on the
specific value of disorder used. Another observation of Ref.\ 
\cite{WWP98} is the duality in $U$ and $\sqrt{24}/U$ for very large
$|U|$ (small $1/|U|$). The data in Fig.\ \ref{fig-tipdm-l2_u} are only
compatible with this duality for the large disorder $W=5$. For the
smaller disorders and for the range of interactions shown, we do not
observe the duality. We emphasize that this may be due to restricting
ourselves to values $U \leq 4$.

For $E \neq 0$, the independence of the enhancement on the sign of the
interaction $U$ is no longer valid. In Fig.\ \ref{fig-tipdm-l2_u_ew}
we show $\lambda_2(U)/\lambda_2(0)$ for the same disorders as before,
but now at energies $E= \pm 1$. We find that the enhancement for $U=1$
is larger at $E=1$ than at $E=-1$, whereas exactly the opposite is
true for $U=-1$.  Thus we see that for positive (negative) $U$ the
energies of TIP states are shifted towards higher (lower) values,
eventually leading to the formation of the aforementioned Hubbard
bands. In Fig.\ \ref{fig-tipdm-l2_u_e} we show the localisation
lengths at several values of $E$ for $W=4$.  As expected from the
discussion above the localisation lengths are always smaller than at
the band center. The enhancements, however, which are shown in Fig.\ 
\ref{fig-tipdm-l2_u_ew}, can be equally large for $E=0$ and $E\neq 0$.

%
%

\section{FSS applied to the TIP problem}
\label{sec-tipdm-fss}

In order to overcome the problems with the finite chain lengths, we
now proceed to use the FSS technique and construct FSS curves for each
$U= 0, 0.1, \ldots, 1$.  In Fig.\ \ref{fig-tipdm-l2_m} we show the
data for the reduced localisation lengths $\lambda_2/M$ which is to be
rescaled just as in the standard TMM \cite{MK83} to obtain the
localisation length $\xi_2$ for the infinite system. Note that data
for small $W$ is rather noisy and will thus most likely not give very
accurate scaling. Furthermore, in Fig.\ \ref{fig-tipdm-l2_um} we show
$\lambda_2$ for $W=3$ and $W=9$ and all interaction strengths $U= 0,
0.1, \ldots 1.0$. We see that whereas for $W=3$ the values of
$\lambda_2$ for $U=0$ show only small variations for large $M$, the
$W=9$ data shows a rapid increase of $\lambda_2$ as $M$ increases.
This is due to the numerical problem of estimating a small
localisation length of the order of 1 in a large system by Eq.\ 
(\ref{eq-lambda2}). It is most pronounced for small $U$ where the
localisation lengths are the smallest. Going back to Fig.\ 
\ref{fig-tipdm-l2_m}, we see that this does not influence the reduced
localisation lengths $\lambda_2/M$ very much and thus is not expected
to deteriorate the FSS procedure. However, in order to set an absolute
scale in the FSS procedure, one usually fits the smallest localisation
lengths of the largest systems to $\lambda_2/M = \xi_2/M + b
(\xi_2/M)^2$ with $b$ small \cite{MK83}.  In the present case this
would mean taking the unreliable data for $W=9$.  Therefore, for each
$U$ we fit to the localisation length at $W=3$ and adjust the absolute
scale of $\xi_2$ accordingly.  In Fig.\ \ref{fig-tipdm-fss} we show
the resulting scaling curves $\lambda_2/M = f(\xi_2 /M)$ for $U=0$,
$0.2$ and $1.0$.  Note that, as expected from Fig.\ 
\ref{fig-tipdm-l2_m}, FSS is not very accurate for small $W$. The
previously discussed unreliable data for large $W$ are visible only in
very small upward deviations from the expected $1/M$ behavior. In
Fig.\ \ref{fig-tipdm-xi2_w} we show the scaling parameters $\xi_2$
obtained from the FSS curves of Fig.\ \ref{fig-tipdm-fss}.

A simple power-law fit $\xi_2 \propto W^{-2\alpha}$ in the disorder
range $W\in [1,5]$ yields an exponent $\alpha$ which increases with
increasing $U$ as shown in the inset of Fig.\ \ref{fig-tipdm-xi2_w},
e.g., $\alpha= 1.1$ for $U=0$ and $\alpha= 1.55$ for $U=1$.  Thus,
although in Fig.\ \ref{fig-tipdm-l2_w} the $\lambda_2$ data at $M=200$
nicely follows $\lambda_1/2$ for $U=0$, we nevertheless find that
after FSS with data from all system sizes, $\xi_2(0)$ still gives a
slight enhancement.  Because of this in the following we will compare
$\xi_2(U>0)$ with $\xi_2(0)$ when trying to identify an enhancement of
the localisation lengths due to interaction. We emphasize that the
slight dip in the $\alpha(U)$ curve around $U= 0.7$ has also been
observed in Ref.\ \cite{SK97}.
 
The derivation of Eq.\ (\ref{eq-shep}) is based on a mapping of the
TIP Hamiltonian onto an effective random matrix model while assuming
uncorrelated interaction matrix elements \cite{S94}.  In Refs.\ 
\cite{VRS97} and \cite{PS97} a more accurate estimate of the matrix
elements of the interaction in the basis of SP states was calculated
showing that the original estimates of Ref.\ \cite{S94} were
oversimplified. The authors of Ref.\ \cite{PS97} then considered a
more appropriate effective random matrix model and obtained
$\lambda_2\,\propto \, \lambda_1^{\beta}$ for large values of
$\lambda_1$.  To correct for smaller values of $\lambda_1$ they
suggested a more accurate expression should be $\lambda_2\, \propto\,
\lambda_1^{\beta}(1+c/\lambda_1)$. An important prediction of this
work is that $\beta$ is $U$-dependent with $\beta$ ranging from $1$ at
small $U$ and very large $U$ to nearly $2$ for intermediate values $U
\propto t$. Using our data obtained from FSS, we translate this fit
function into
\begin{equation}
  \xi_2(U) \propto \xi_2(0)^{\beta} \left( 1 + \frac{c}{\xi_2(0)} \right).
\label{eq-ps}
\end{equation}
We remark that the actual least-squares fit is performed with the
numerically more stable fit function $y = a+\beta x+c*\exp(-x)$ with
$y=\ln[\xi_2(U)]$ and $x=\ln[\xi_2(0)]$.  In Fig.\ \ref{fig-tipdm-ps}
we show results for disorders $W \in [1,6]$ and various $U$. As can be
seen easily, the fit is rather good and does indeed capture the
deviations from a simple power-law $\xi_2(U) \propto \xi_2(0)^\beta$
for small localisation lengths. In the inset of Fig.\ 
\ref{fig-tipdm-ps} we show the variations of $\beta$ with $U$ for both
the simple power-law and the fit according to Eq.\ (\ref{eq-ps}). We
note that contrary to Ref.\ \cite{PS97}, we find $\beta < 1.5$ for all
$U$ values considered.

In Ref.\ \cite{OWM96} is has been suggested that a more suitable
functional dependence of the TIP localisation lengths is given by
$\lambda_2 = \lambda_1/2 + c |U| \lambda_1^2$. Using the $\xi_2$ data
and taking instead of $\lambda_1/2$ the more suitable $\xi_2(0)$ we
translate this proposed fit as
\begin{equation}
  \xi_2(U) - \xi_2(0) \propto \xi_2(0)^\beta.
\label{eq-oppen}
\end{equation}
In Fig.\ \ref{fig-tipdm-xi2_0} we plot $\xi_2(U)-\xi_2(0)$ vs.\ 
$\xi_2(0)$ for $U$-values $0.2$ and $1$. We find that instead of being
able to fit the data with a single $\beta$, it appears that for small
$\xi_2(0)<10$ we have $\beta\approx 2$, whereas for larger $\xi_2(0)$
we find $\beta\approx 3/2$. Note that a crossover from the functional
form (\ref{eq-oppen}) with $\beta=2$ to $\beta=3/2$ has been suggested
previously \cite{WP96}. However, in that work, the exponent $3/2$ is
supposed to be relevant for larger disorders, opposite to what we see
here. As pointed out previously, our FSS may give rise to artificially
small values of $\xi_2(U)$ close to the largest system size, and one
might want to argue that the reduction in slope is due to this effect.
However, we emphasize that the crossover observed in Fig.\ 
\ref{fig-tipdm-xi2_0} occurs at $W=2.5$ where FSS appears to be still
reliable. We remark that an exponent close to $1.5$ for small $W$ has
also been found in Ref.\ \cite{WP97} from a multifractal analysis.

The most recent suggestion of how to describe the TIP localisation
data is due to Song and Kim \cite{SK97}. They assume a scaling form 
\begin{equation}
  \xi_2 = W^{-\beta_0} g(|U|/W^{\Delta})
\label{eq-sk}
\end{equation}
with $g$ a scaling function and obtain $\Delta = 4$ by fitting the
data. Choosing the same value for $\Delta$ we find that our data can
be best described when $\beta_0$ is related to the disorder dependence
of $\xi_2$ as $(\beta - \beta_0)/\Delta \approx 1/4$. However, the
scaling is only good for $W\in [1,5]$ and $U \geq 0.3$. Unfortunately,
even using our varying exponent $\beta(U)$, we have not been able to
obtain a good fit to the scaling function with the data for all $U$.
We emphasize that the $\xi_2$ values for $U \leq 0.2$ are smaller than
for $U\geq 0.3$ and thus numerically quite reliable. 

A much better scaling can be obtained when plotting
\begin{equation}
  \xi_2(U) - \xi_2(0) = \tilde{g}\left[ f(U) \xi_2(0) \right]
\label{eq-vo}
\end{equation}
with $f(U)$ determined by FSS.  In Fig.\ \ref{fig-tipdm-vo} we show
the resulting scaling curves and scaling parameters $f(U)$. Note that
the scaling is valid for $U= 0.1, 0.2, \ldots 1.0$ and most disorders
$W\in [0.6,9]$. Again we see the crossover from a slope 2 to a slope
3/2. Deviations from scaling occur for large and very small values of
$\xi_2(U)$ and are most likely due to numerical inaccuracy as
discussed before. The behavior of $f(U)$ as shown in the inset
indicates that for $U \geq 0.3$ a linear behavior $f(U) \propto U$ may
be valid which then translates into $U^2$ ($U^{3/2}$) dependence of
$\xi_2(U) - \xi_2(0)$ in the regions of Fig.\ \ref{fig-tipdm-vo} with
slope 2 (3/2). However, for $U \leq 0.5$, one could also argue that
$f(U)\propto \sqrt{U}$ which would give $\xi_2(U) - \xi_2(0) \propto
U$ ($U^{3/4}$) in these regions. We note that a crossover from $U$ to
$U^2$ behavior had been proposed in Ref.\ \cite{WP96}, but it should
appear at larger values of $U$ and also be $W$ dependent.  We observe
that the best fit to the $f(U)$ data is obtained by a logarithmic
$U$-dependence as indicated in the inset.

Thus in summary it appears that our data cannot be described by a
simple power-law behavior with a single exponent neither as function
of $W$, nor as function of $\xi_2(0)$, nor after scaling the data onto
a single scaling curve.  The best power-law fit is obtained in Fig.\ 
\ref{fig-tipdm-ps} with an exponent $\beta(U)$, whereas after scaling
of $\xi_2(U) - \xi_2(0)$ onto a single curve we need at least two
powers to describe the scaling curve as shown in Fig.\ 
\ref{fig-tipdm-vo}. Lacking a convincing explanation as to what fit
function should be correct, we must at present be content with letting
the reader decide for himself.

%
%

\section{The interacting electron-hole problem}
\label{sec-iehdm}

Let us now consider what happens when the two particles are in
different random potentials such that in general $\epsilon^1_n \neq
\epsilon^2_n$.  Such a problem is relevant for the proposed
experimental verification of the TIP effect by optical experiments in
semiconductors \cite{G98}.  In these experiments, the electron will be
in a random potential different from that of the hole. Thus this
choice of random potential models the case of interacting
electron-hole pairs (IEH).  Again, we will mostly be concerned with
the case of repulsive interactions. In the experimental situation, of
course, the interaction is attractive. As shown in Fig.\ 
\ref{fig-iehdm-l2_u} we again have $\lambda_2(-U) = \lambda_2(U)$ for
$E=0$ and thus our results apply also to the case $U<0$. For
simplicity, we also take the width of the disorder distribution to be
the same for both particles.

As for TIP we compute the IEH localisation lengths by the DM along the
diagonal. Comparing with the results presented in the previous
sections, we find that the results for IEH are very similar to the
case of TIP.  FSS is possible and again the best fit is obtained by
using Eq.\ (\ref{eq-ps}) as shown in Fig.\ \ref{fig-iehdm-ps} for
$U=0$, $0.1$, $0.2$, $0.5$ and $1.0$. The values of the power
$\beta(U)$ shown in the inset of Fig.\ \ref{fig-iehdm-ps} are also
much as before.  Thus we can conclude that the case of IEH is very
close to the TIP problem.

%
%

\section{The 2D Anderson model with an additional diagonal potential}
\label{sec-ab2ddm}

In Ref.\ \cite{VRS97}, two of us argued that straightforward
application of the random matrix models (RMM) \cite{S94} and the
block-scaling picture (BSP) \cite{I95} gives rise to an erroneous
enhancement of the SP localisation length $\xi_1$ in a 2D Anderson
model with additional random perturbing potential $U(n) \in [-U,U]$
along the diagonal.  In fact, the same is true if the potential along
the diagonal is taken to be constant, i.e. $U(n)= U$. Although it
appears obvious that no such SP enhancement should exist, we have
checked it here with the DM. In Fig.\ \ref{fig-a2ddm-xi_u} we show
examples of the resulting SP localisation lengths $\xi_1$ obtained as
before from FSS of SP localisation lengths $\lambda_1$ calculated for
various system sizes $M=51, \ldots, 261$, disorders $W$ and potentials
$U= 0, 0.1, \ldots, 1$.  As expected, we find that for large disorders
$W>5$, the data is well described by the 2D TMM results already
presented in section \ref{sec-test}. There are only small changes due
to the additional random potential, all of which tend to decrease the
localisation lengths as they should.
%
%
This is in contrast to the straightforward application of the RMM and
the BSP \cite{VRS97} which therefore fail for the 2D SP Anderson model
with additional random potential along the diagonal. Of course this
does not mean that these methods also have to fail for TIP, where, as
we have shown in the previous sections, a tendency towards
delocalisation due to interaction definitely exists.

%
%

\section{Conclusions}
\label{sec-concl}

In conclusion, we have presented detailed results for the localization
lengths of pair states of two interacting particles in 1D random
potentials. By using the DM to calculate the Green function along the
diagonal it is possible to consider the 2D Anderson model and the
problem of two interacting particles in 1D within the same numerical
formalism. We have checked that for the 2D Anderson model without
interaction the infinite system size results obtained via FSS from the
DM data are in good agreement with results obtained from the standard
TMM especially for localisation lengths up to the largest system sizes
we have considered. It is also apparent that the DM data deviate from
the TMM only towards smaller localisation lengths and hence no
artificial enhancement of localisation lengths due to the DM approach
is expected.

For TIP in 1D we observe an enhancement of the two-particle
localisation length up to $75\%$ due to onsite interaction. This
enhancement persists, unlike for TMM, in the limit of large system
size and after constructing infinite-sample-size estimates from the
FSS curves.  We have tried to fit our results to various suggested
models. The best fit was obtained with Eq.\ (\ref{eq-ps}) in which the
enhancement $\xi_2(U)/\xi_2(0)$ depends on an exponent $\beta$ which
is a function of the interaction strength $U$. Such a $U$-dependent
exponent had been previously predicted in Ref.\ \cite{PS97} for
interaction strengths up to $U=1$ with $\beta$ up to 2.  However, we
find that $\beta$ reaches at most $1.5$ for $U=1$. Thus we do not see
a behavior as in Eq.\ (\ref{eq-shep}) with exponent $2$ when using the
fit function of Ref.\ \cite{PS97}.  On the other hand, after scaling
the data onto a single scaling curve and using the fit function
(\ref{eq-oppen}) as proposed with $\beta=2$ in Ref.\ \cite{OWM96}, we
find indeed $\beta=2$ for not too small disorder strength, e.g., $W
\geq 2.5$ for $U=1$, but observe a crossover to a behavior with
$\beta=3/2$ for smaller $W$.  For values of $U \gtrsim 1.5$ we observe
that the enhancement decreases again; the position of the maximum
depends upon $W$.  Very similar results are produced by placing the
two particles in different potentials which is of relevance for a
proposed experimental test of the TIP effect \cite{G98}.

As a final check on our results we consider the effect of an
additional on-site potential (both random and uniform) on the results
for the SP 2D Anderson model. As one may expect for the case of an
additional random potential one observes only a small decrease in the
localisation length while for an additional uniform potential there is
a small change in $\xi_1$ towards decreasing localization lengths for
positive $U$.

\acknowledgments

We thank E.\ McCann, J.\ E.\ Golub, O.\ Halfpap, S. Kettemann and D.\ 
Weinmann for useful discussions.  R.A.R.\ gratefully acknowledges
support by the Deutsche Forschungsgemeinschaft through SFB 393.

%
%

%
%
\newpage
\newcommand{\figwidth}{0.9\columnwidth}
\newcommand{\figheight}{0.9\textheight}
\newcommand{\figheightb}{0.6\textheight}



\clearpage
\begin{figure}[th]
\centerline{\epsfig{figure=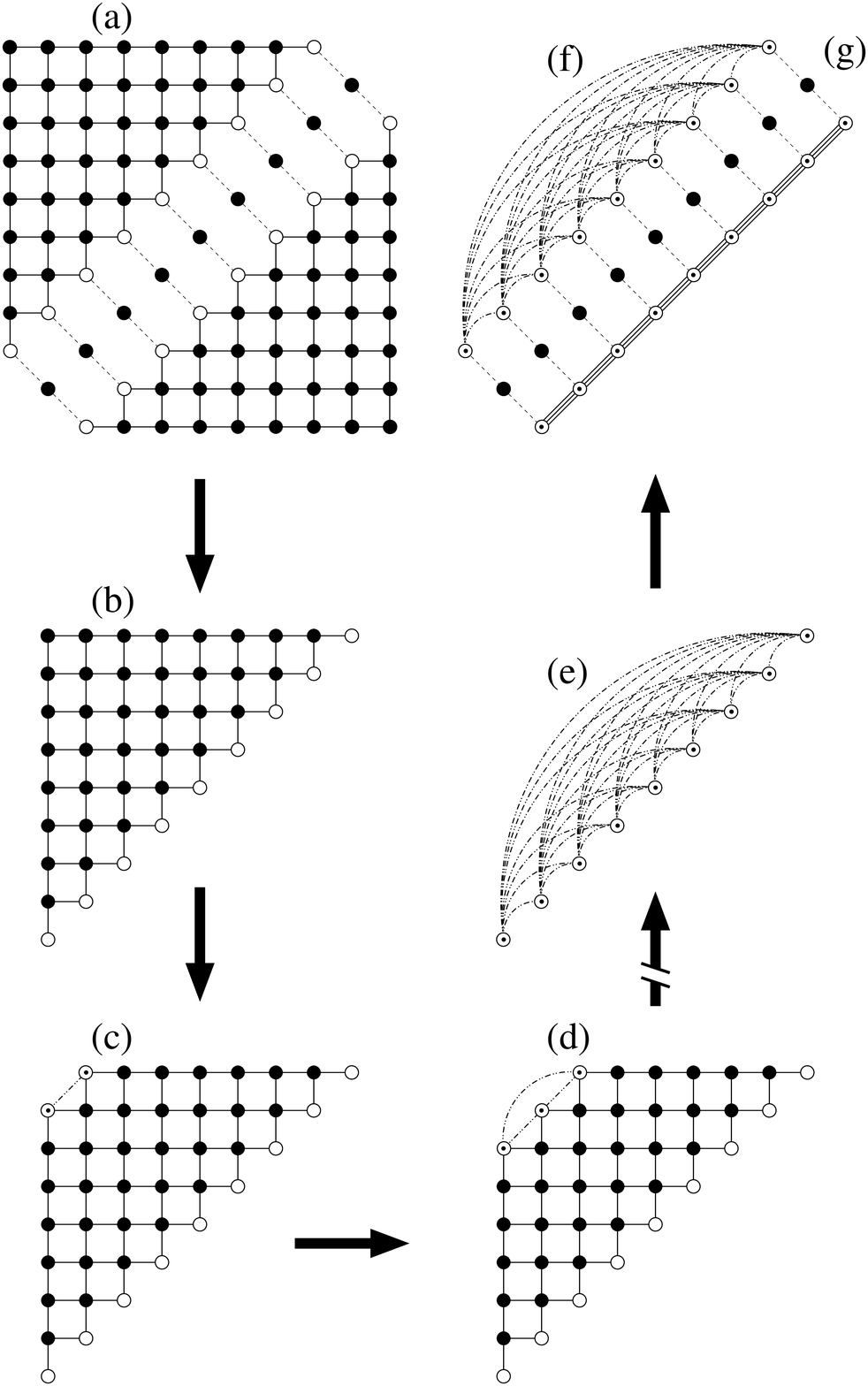,height=\figheightb} } 
\caption{\label{fig-dm}
  The decimation process: (a) The lattice is split into two parts.
  (b) Each half is then `decimated' independently. (c,d) Sites
  ($\bullet$) and nearest-neighbor hops ($-$) in the original lattice
  are replaced successively by effective long-range hops ($- \cdot
  \cdot -$) between the effective sites ($\odot$).  (e) This
  decimation continues until (f) the diagonal is reached. (g) Finally,
  the two halfs are recombined.}
\end{figure}


\clearpage
\begin{figure}[th]
\centerline{\epsfig{figure=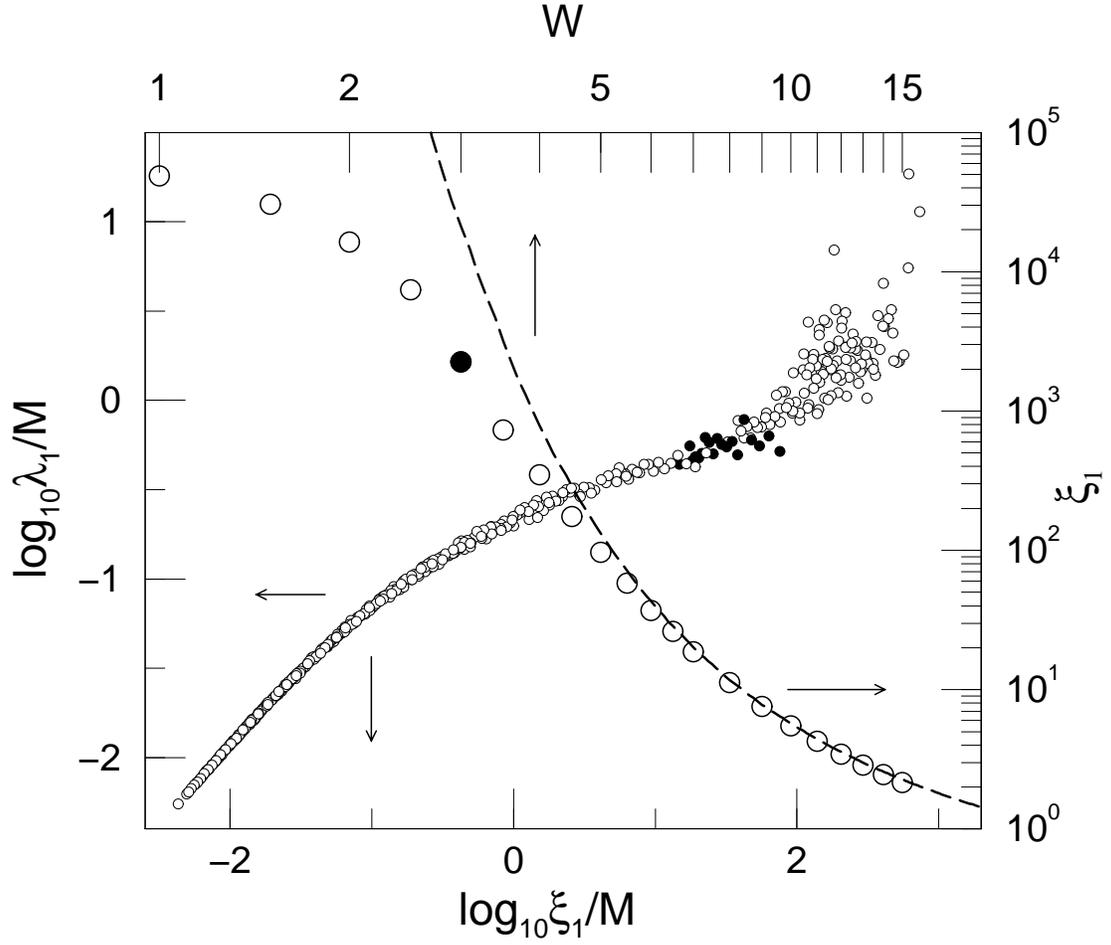,width=\figwidth} }
\caption{\label{fig-a2ddm-xi}
%
  Small symbols (with left and bottom axis) denote the FSS curve used
  to compute the $\xi_1$ values. Large symbols (with right and top
  axis) indicate SP localisation lengths $\protect\sqrt{2}\ \xi_1$
  obtained by DM along the diagonal ($\circ$) and $\xi_1$ computed by
  TMM of quasi-1D strips (dashed line) after FSS. The filled symbols
  correspond to a disorder at which FSS appears to be unreliable.}
\end{figure}


\clearpage
\begin{figure}[th]
\centerline{\epsfig{figure=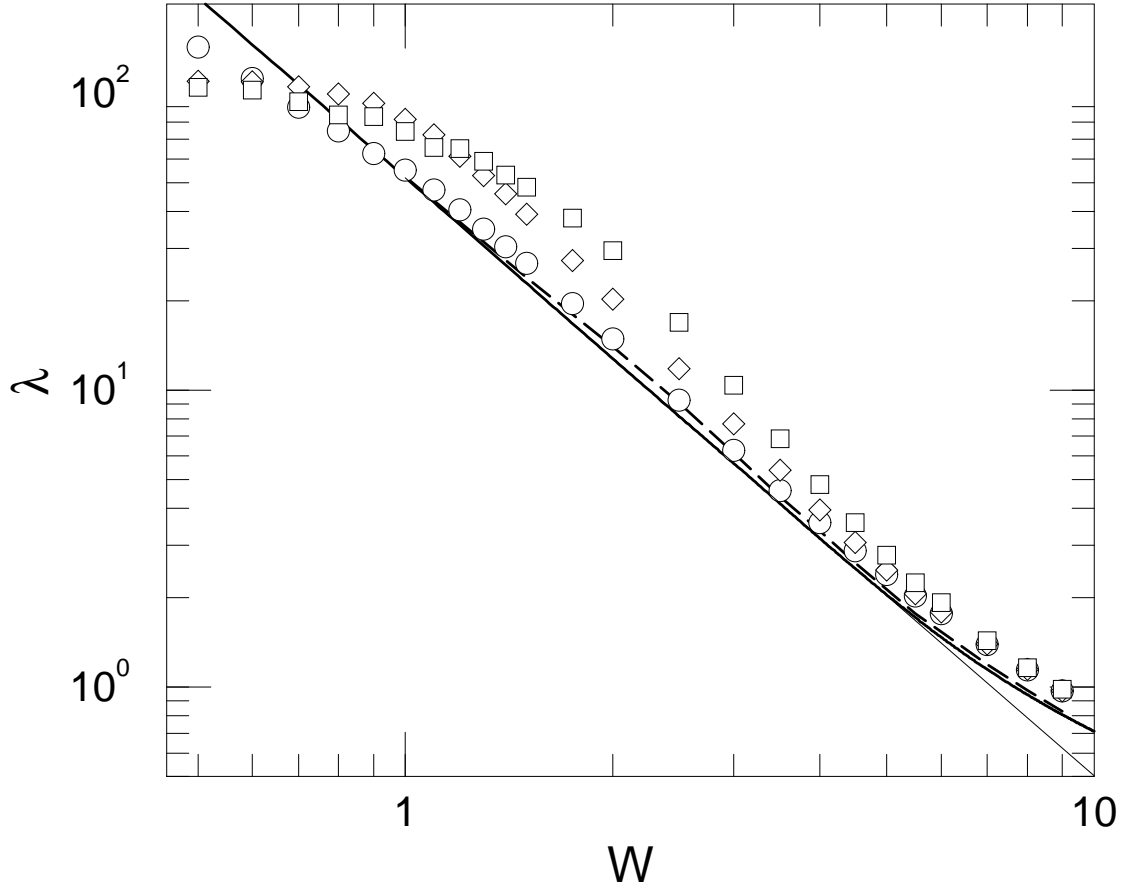,width=\figwidth} }
\caption{\label{fig-tipdm-l2_w}
%
  Two-particle localisation length $\lambda_2$ at energy $E=0$ for
  system size $M=201$ and interaction strength $U=0$ ($\bigcirc$),
  $U=0.2$ ($\Diamond$) and $U=1$ ($\Box$). The thick solid line
  represents 1D TMM data for SP localisation length $\lambda_1/2$, the
  dashed line is computed from the convolution of SP Green functions
  in Eq.\ (\protect\ref{eq-g1g1}). The thin line is the perturbative
  result $\lambda_1/2 \approx 52/W^2$. }
\end{figure}

\clearpage
\begin{figure}[th]
\centerline{\epsfig{figure=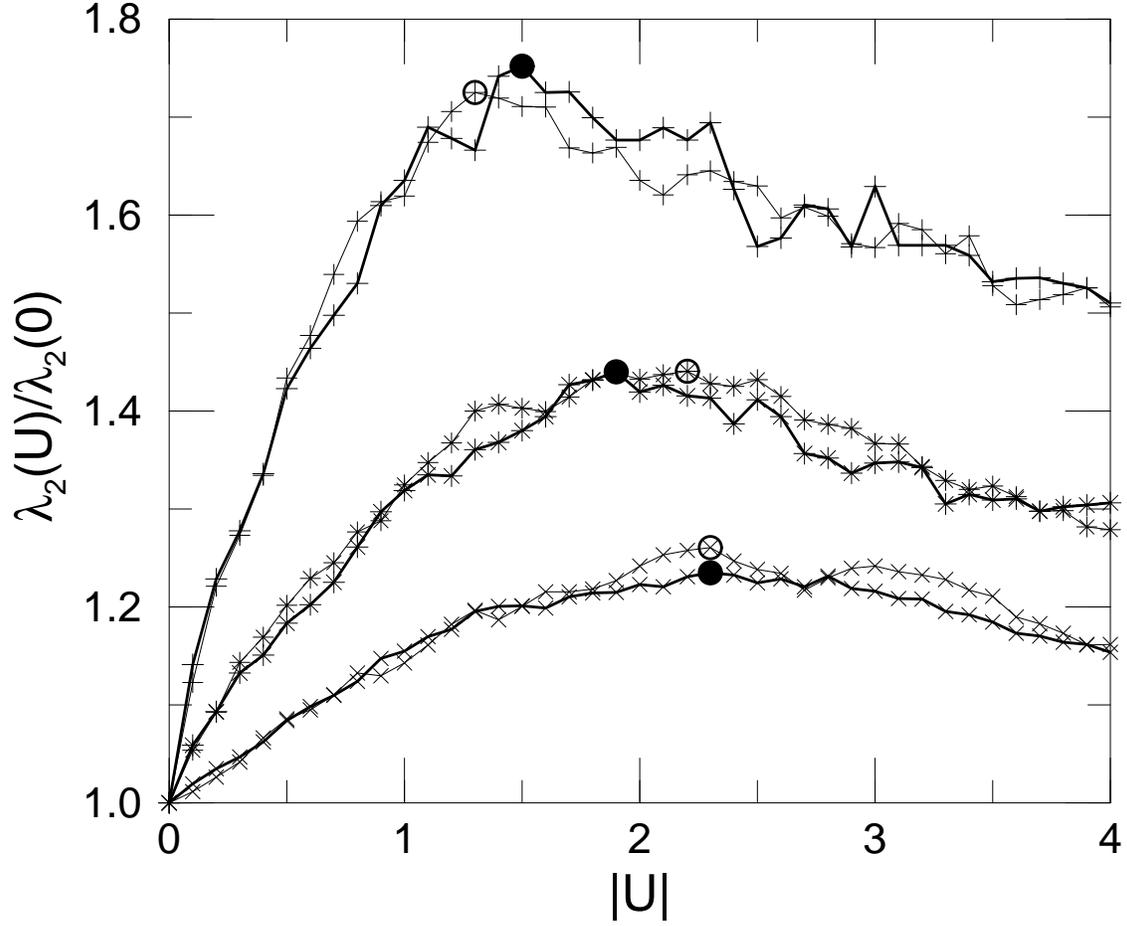,width=\figwidth} }
\caption{\label{fig-tipdm-l2_u}
%
  Enhancement $\lambda_2(U)/\lambda_2(0)$ for TIP as a function of
  interaction strength $U$ at $E=0$ for disorder $W=3$ ($+$), $W=4$
  ($*$), and $W=5$ ($\times$) and $M=201$. The data are averaged over
  100 samples. The thick (thin) lines indicate data for $U>0$
  ($U<0$), full (open) circles denote the maximum for each disorder. }
\end{figure}

\clearpage
\begin{figure}[th]
\centerline{\epsfig{figure=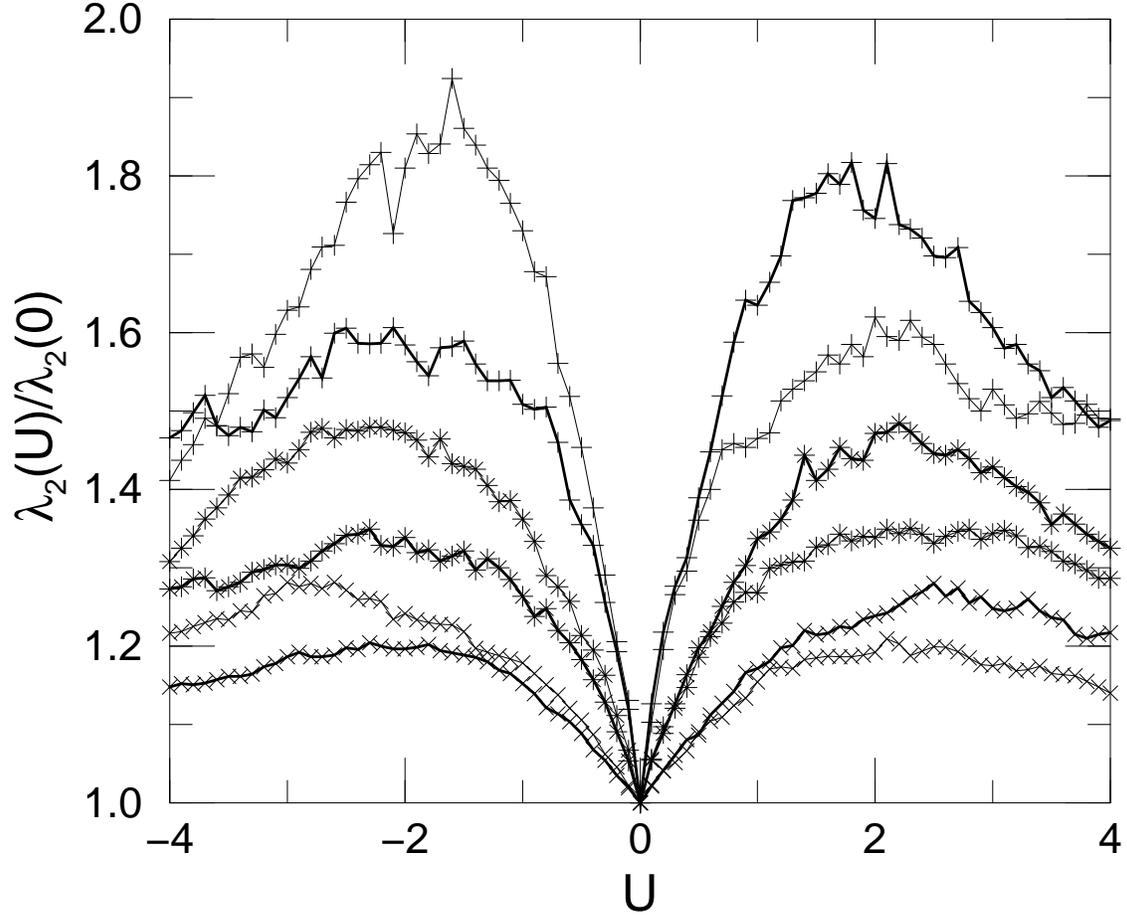,width=\figwidth} }
\caption{\label{fig-tipdm-l2_u_ew}
%
  Enhancement $\lambda_2(U)/\lambda_2(0)$ for TIP as a function of
  interaction strength $U$ at $E=\pm 1$ for disorder $W=3$ ($+$),
  $W=4$ ($*$), and $W=5$ ($\times$) and $M=201$. The thick (thin)
  lines indicate data for $E=1$ ($E=-1$). }
\end{figure}

\clearpage
\begin{figure}[th]
\centerline{\epsfig{figure=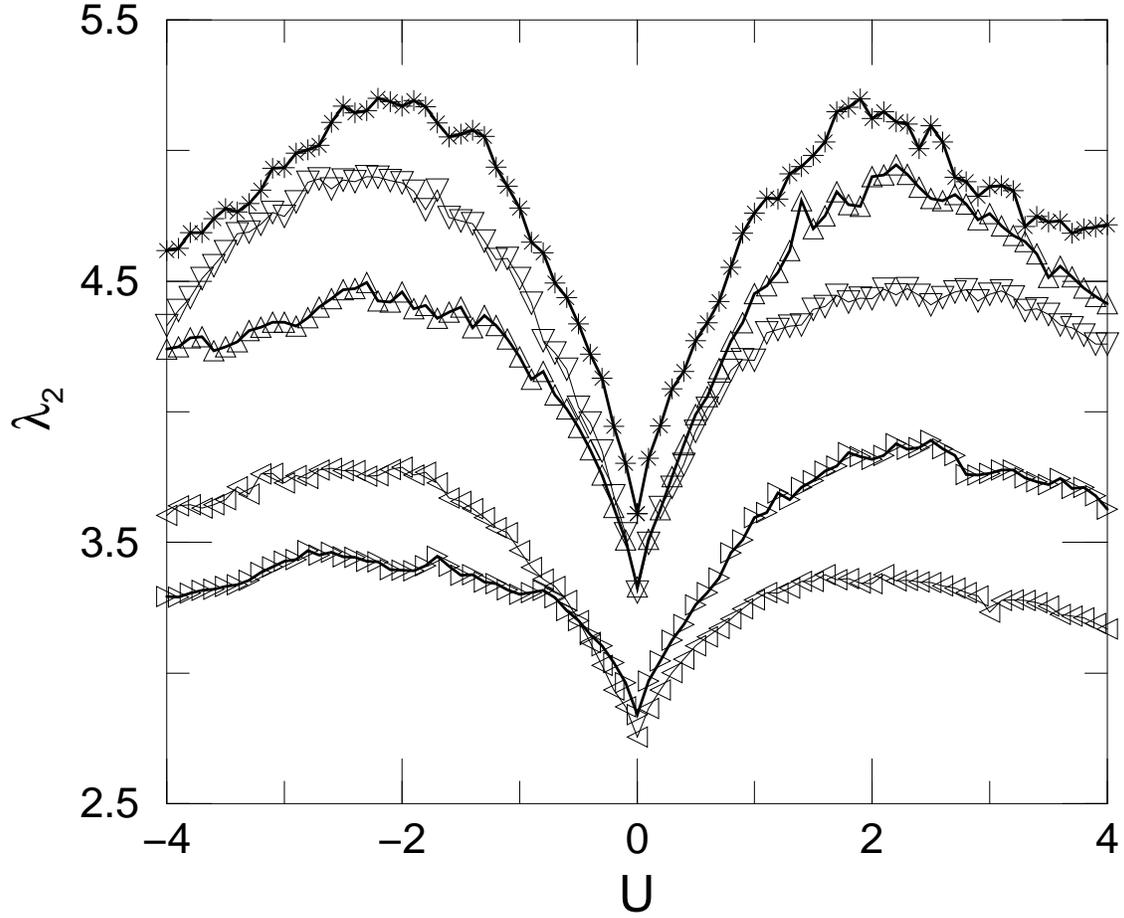,width=\figwidth} }
\caption{\label{fig-tipdm-l2_u_e}
%
  TIP localisation length $\lambda_2$ as a function of interaction
  strength $U$ at $W=4$ for $E=-2$ ($\triangleleft$), $E=-1$
  ($\triangledown$), $E=0$ ($*$), $E=1$ ($\triangle$), and $E=2$
  ($\triangleright$) and $M=201$. The thick (thin) lines indicate
  data for $E\geq 0$ ($E<0$). }
\end{figure}

\clearpage
\begin{figure}[th]
\centerline{\epsfig{figure=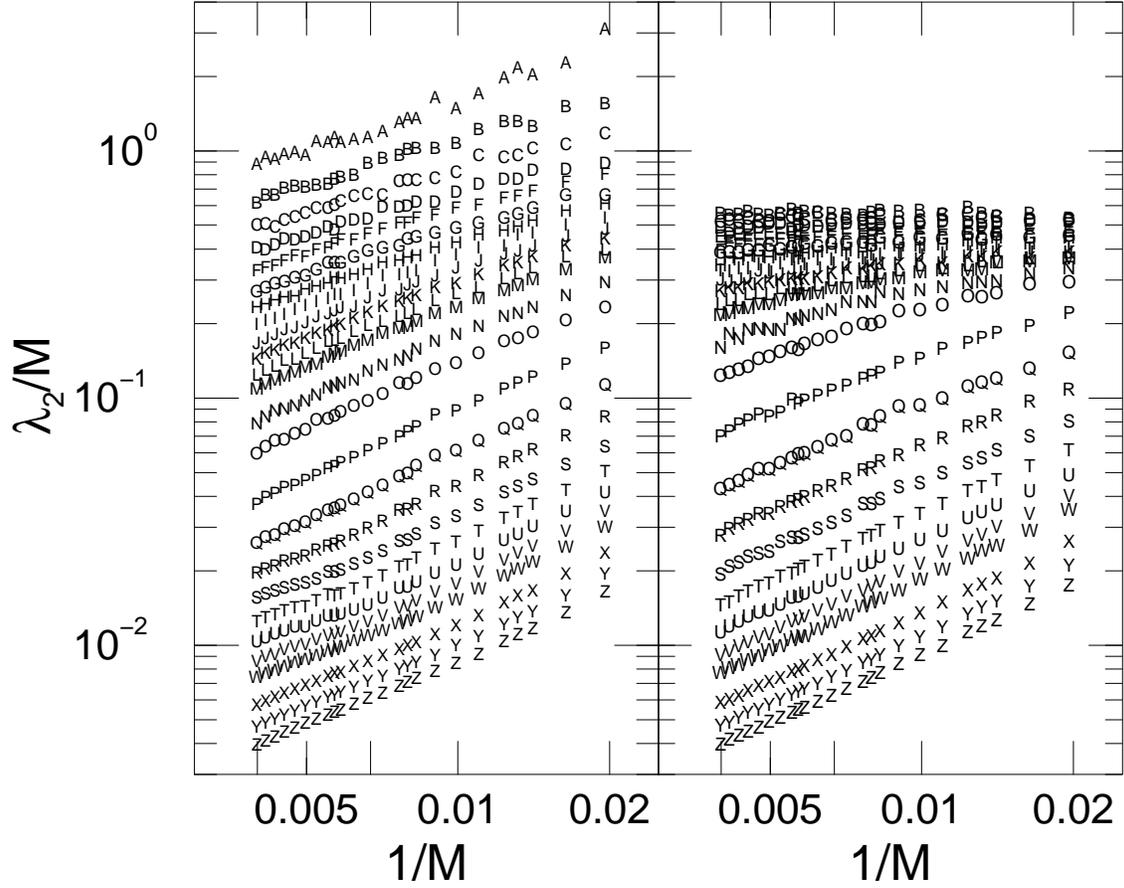,width=\figwidth} }
\caption{\label{fig-tipdm-l2_m}
%
  Reduced TIP localisation lengths $\lambda_2/M$ for $U=0$ (left) and
  $U=1$ (right) for all disorders $W$ and system sizes $M$ obtained by
  averaging 1000 samples for each triple ($U,W,M$). Different letters
  indicate different disorders.}
\end{figure}

\clearpage
\begin{figure}[th]
\centerline{\epsfig{figure=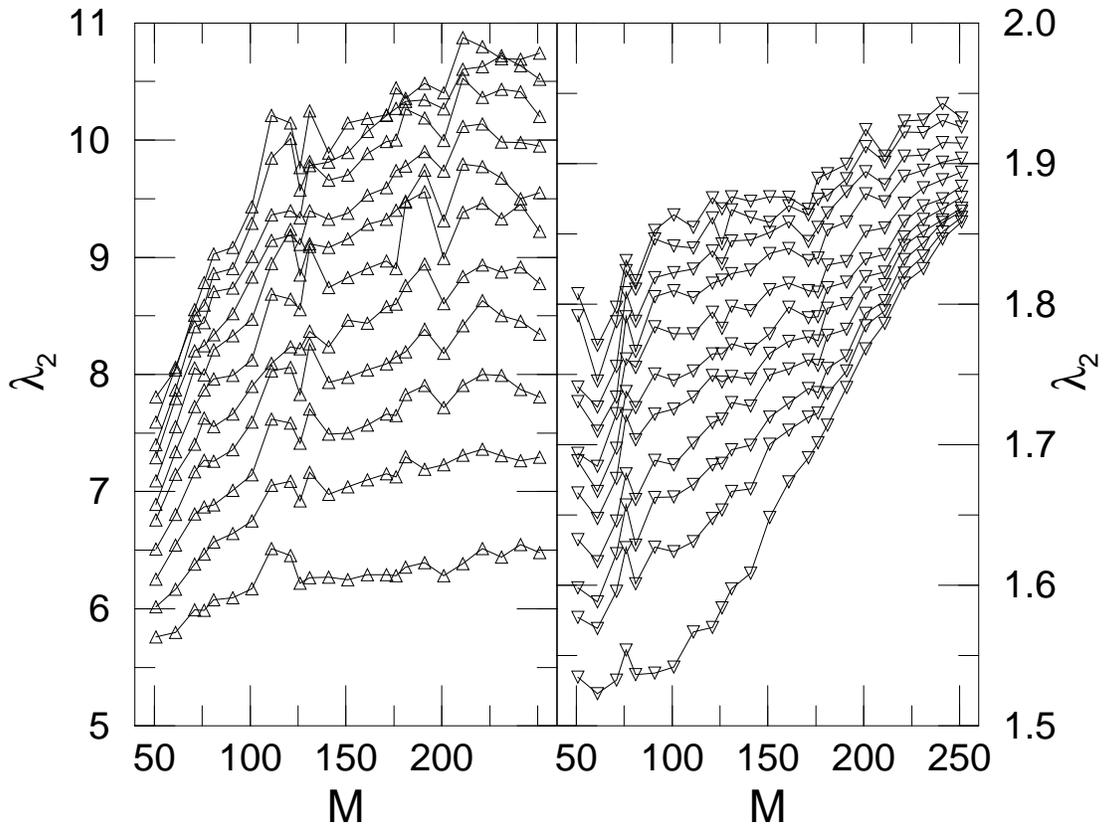,width=\figwidth} }
\caption{\label{fig-tipdm-l2_um}
%
  TIP localisation lengths $\lambda_2$ for $W=3$ (left) and $W=9$
  (right) for $U= 0, 0.1, \ldots, 1$ from bottom to top. We remark
  that we have taken the same set of random numbers for all $U$ to
  increase the numerical efficiency. This is probably the
  reason why for different $U$ the fluctuations in the dependence 
  of $\lambda_2$ on $M$ are similar.}
\end{figure}

\clearpage
\begin{figure}[th]
\centerline{\epsfig{figure=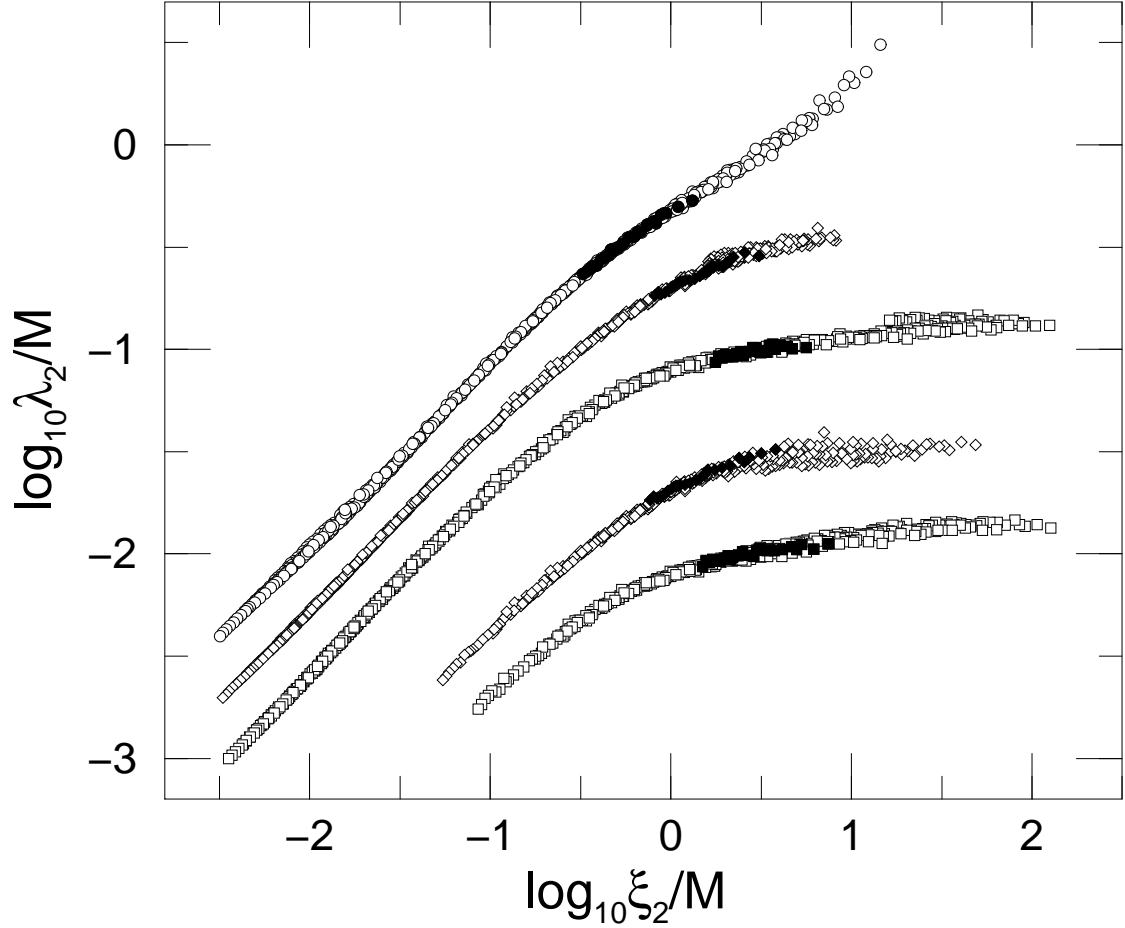,width=\figwidth} }
\caption{\label{fig-tipdm-fss}
%
  Finite-size scaling plot of the reduced TIP localisation lengths
  $\lambda_2/M$ for $U=0$ ($\bigcirc$), $U=0.2$ ($\Diamond$) and $U=1$
  ($\Box$). The data for $U=0.2$ ($U=1$) have been divided by 2 (4)
  for clarity. Data corresponding to $W=1$ are indicated by filled
  symbols. The two curves at the bottom show the data for $U=0.2$ and
  $1$ and $W<2.5$, shifted downward by one order of magnitude for
  clarity, but with the data $W<1$ fitted with scaling parameters 
  obtained from the fit in Fig.\ \protect\ref{fig-tipdm-ps}. }
\end{figure}

\clearpage
\begin{figure}[th]
\centerline{\epsfig{figure=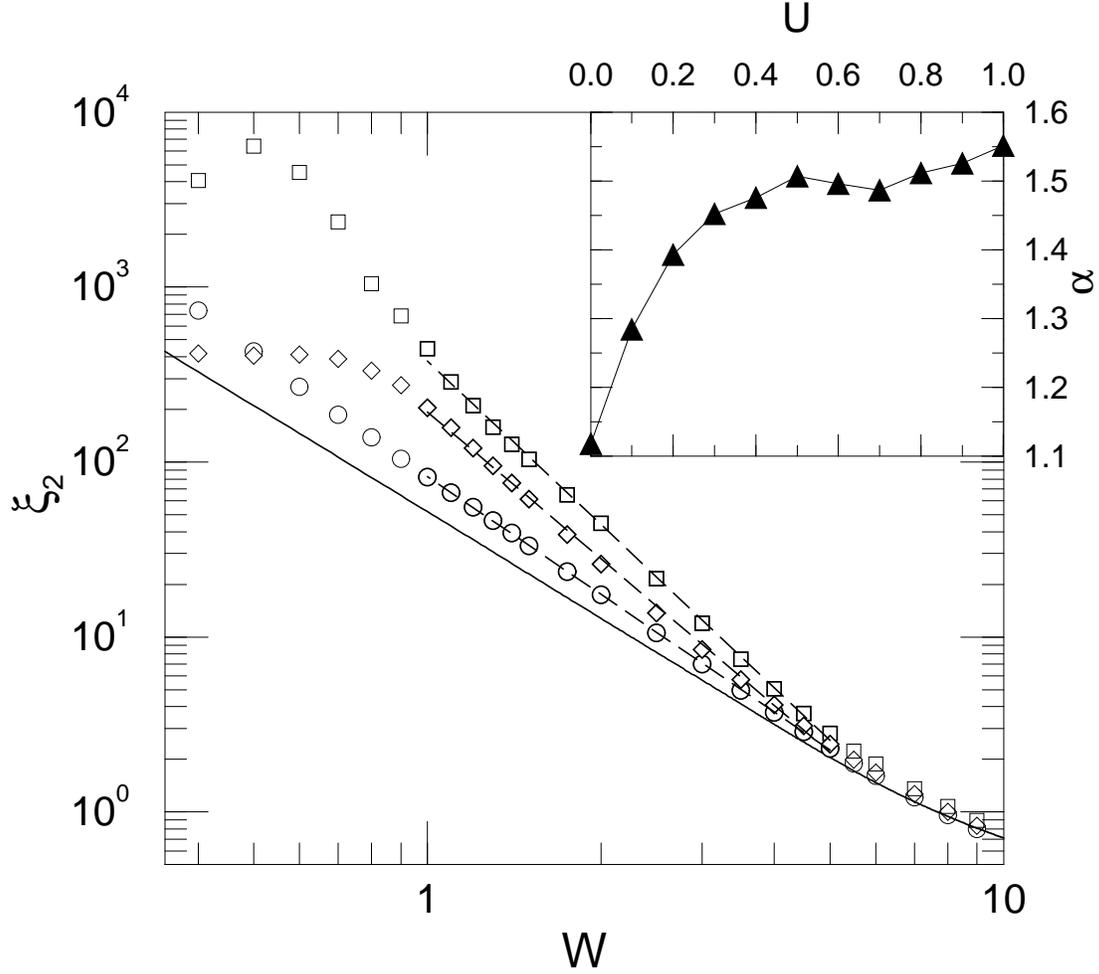,width=\figwidth} }
\caption{\label{fig-tipdm-xi2_w}
%
  TIP localisation lengths $\xi_2$ after FSS for $U=0$ ($\bigcirc$),
  $U=0.2$ ($\Diamond$) and $U=1$ ($\Box$). The solid line represents
  1D TMM data for SP localisation lengths $\lambda_1/2$, the dashed
  lines indicate power-law fits. Inset: Exponent $\alpha$ obtained by
  the fit of $\xi_2 \propto W^{-2\alpha}$ to the data for $U= 0, 0.1, \ldots,
  1$.}
\end{figure}

\clearpage
\begin{figure}[th]
\centerline{\epsfig{figure=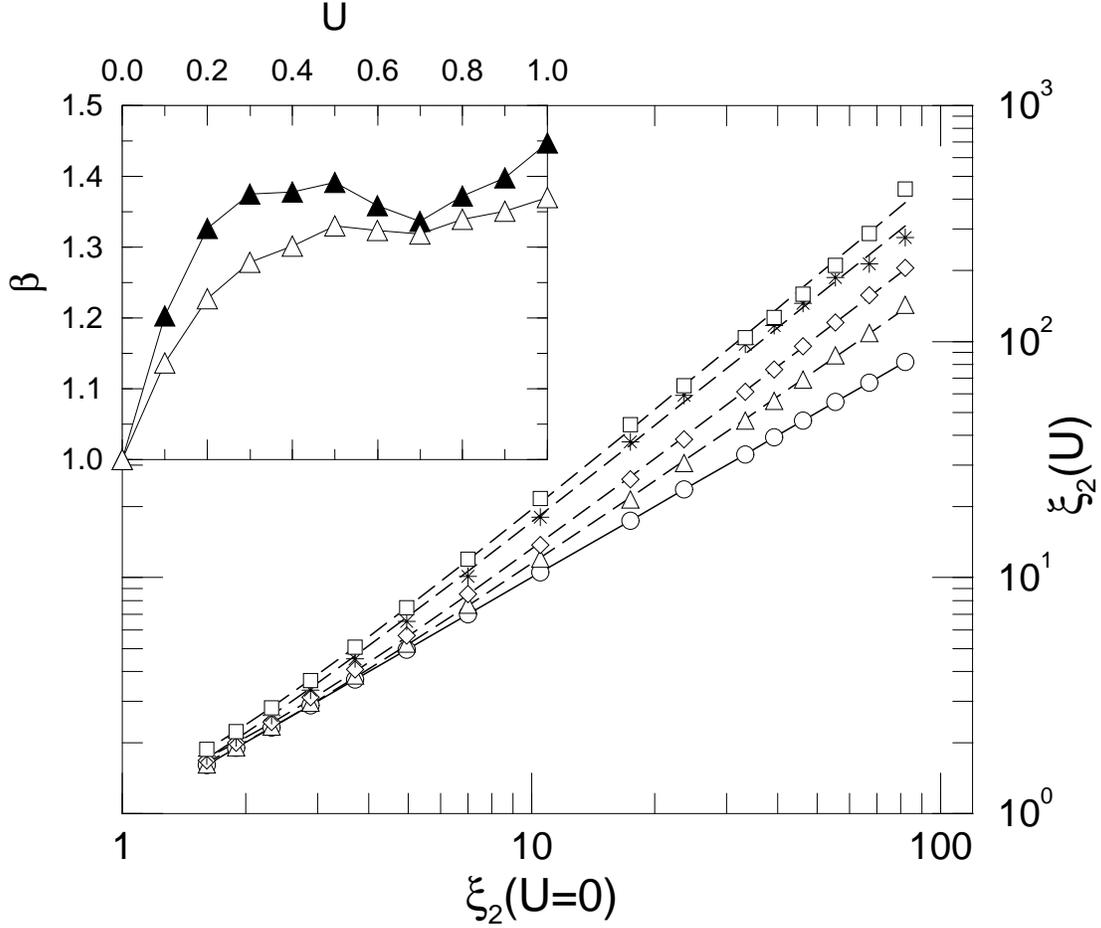,width=\figwidth} }
\caption{\label{fig-tipdm-ps}
%
  TIP localisation lengths $\xi_2(U)$ after FSS for $U=0$
  ($\bigcirc$), $U=0.1$ ($\triangle$), $U=0.2$ ($\Diamond$), $U=0.5$
  ($*$) and $U=1$ ($\Box$) plotted versus $\xi_2(0)$. The data are for
  $W \in [1,6]$. The dashed lines show fits according to Eq.\ 
  (\protect\ref{eq-ps}), the solid line sets the reference for $U=0$.
  Inset: Exponent $\beta$ obtained by the fit of Eq.\ 
  (\protect\ref{eq-ps}) to the data for $U= 0, 0.1, \ldots, 1$. The
  open symbols correspond to the fit with $c = 0$.  }
\end{figure}

\clearpage
\begin{figure}[th]
\centerline{\epsfig{figure=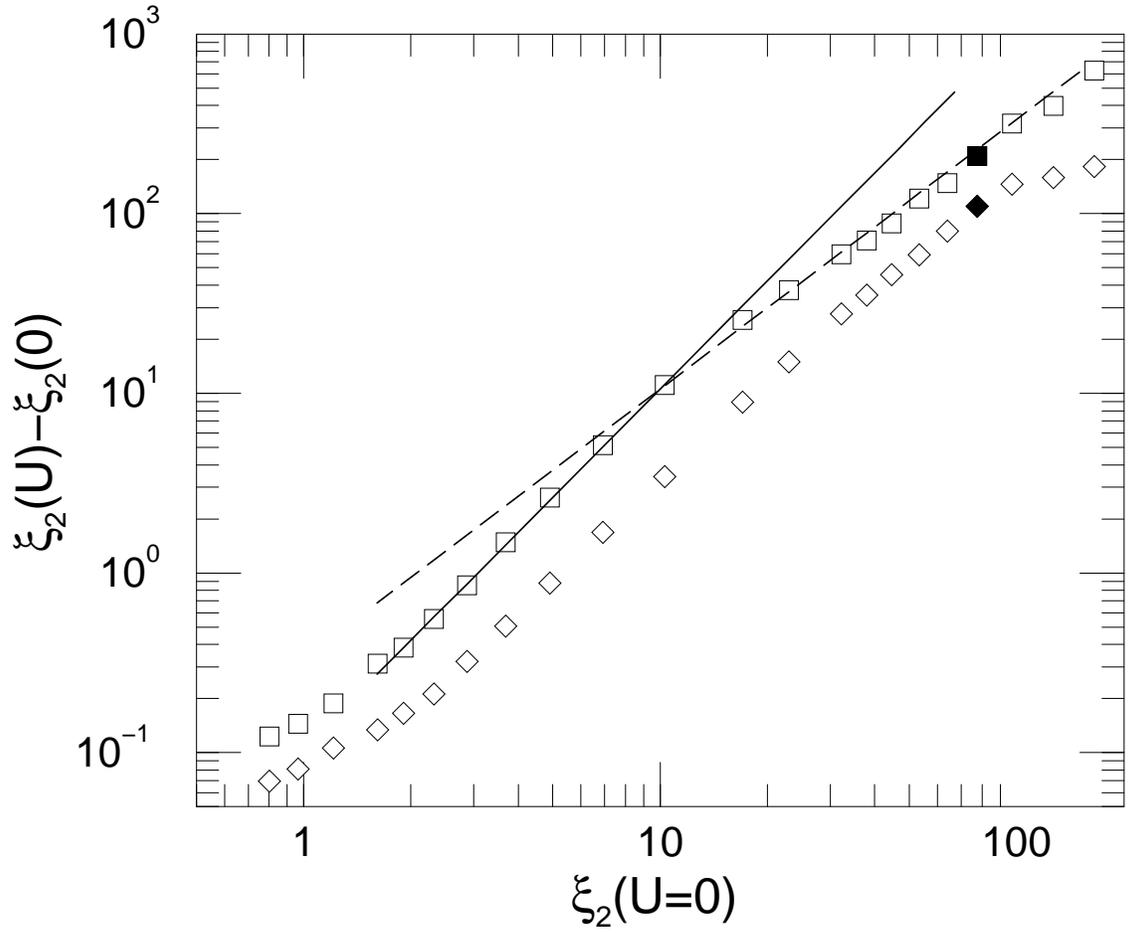,width=\figwidth} }
\caption{\label{fig-tipdm-xi2_0}
%
  TIP localisation lengths $\xi_2$ plotted according to Eq.\ 
  (\protect\ref{eq-oppen}) for $U= 0.2$ ($\Diamond$) and $U=1$
  ($\Box$). The solid line indicates a slope $2$, the dashed line a
  slope $3/2$.  The filled symbols correspond to $W=1$. }
\end{figure}


\clearpage
\begin{figure}[th]
\centerline{\epsfig{figure=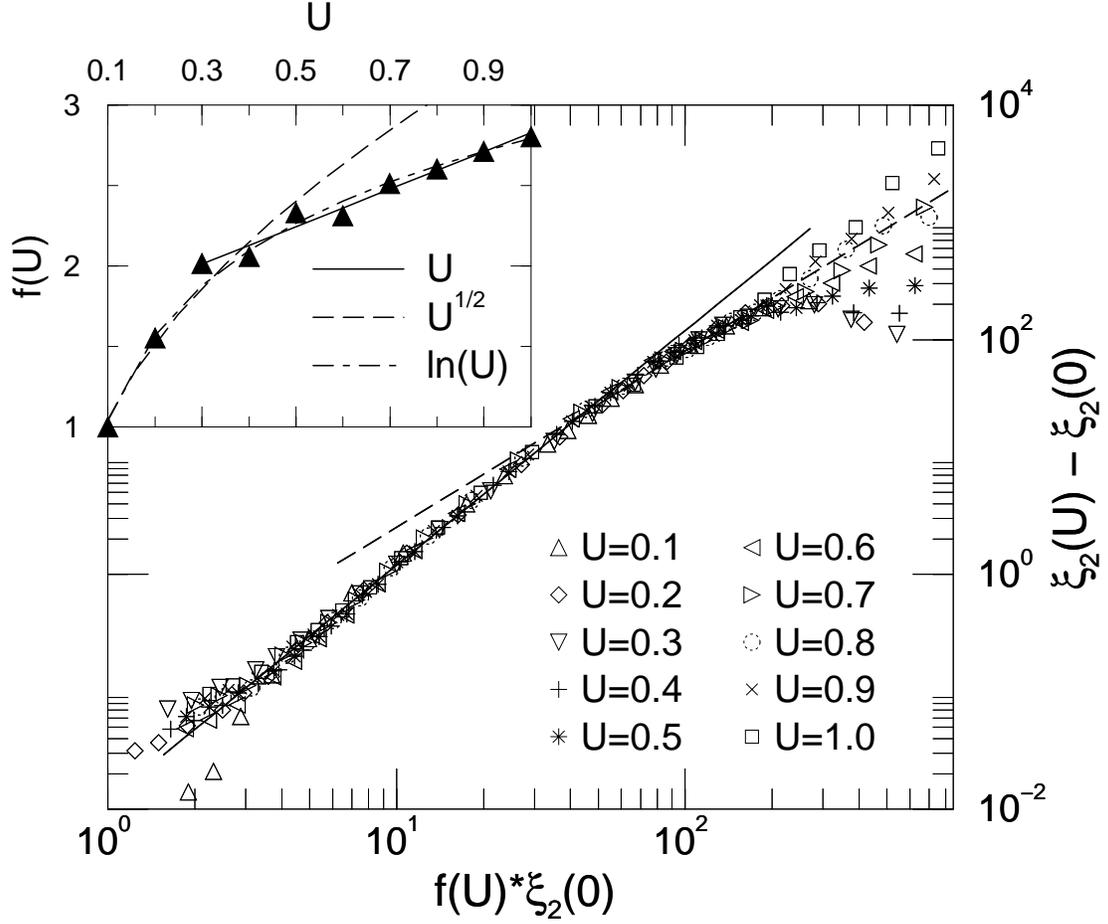,width=\figwidth} }
\caption{\label{fig-tipdm-vo}
%
  Scaling plot of Eq.\ (\protect\ref{eq-vo}) with TIP localisation
  lengths $\xi_2$ for all $U$ and $W\in [0.6,9]$. The solid line
  indicates a slope 2, the dashed line a slope 3/2. Inset: The values
  of $f(U)$ needed to make the data collapse onto the $U=0.1$ curve.
  Solid, dashed and dot-dashed lines are fits of $f(U)$ for $U\geq
  0.3$, $U\leq 0.5$, and all $U$, respectively. }
\end{figure}


%

\clearpage
\begin{figure}[th]
\centerline{\epsfig{figure=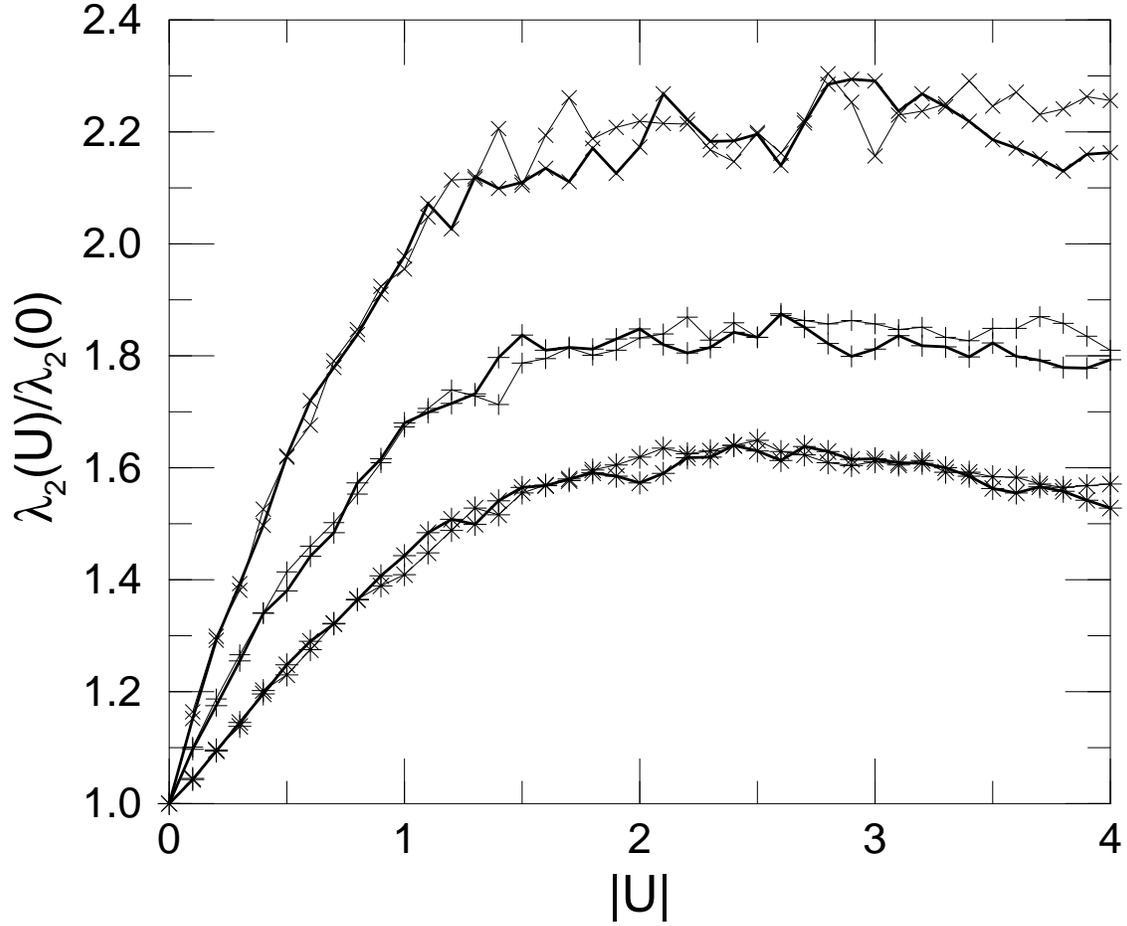,width=\figwidth} }
\caption{\label{fig-iehdm-l2_u}
%
  Enhancement $\lambda_2(U)/\lambda_2(0)$ for IEH as a function of
  interaction strength $U$ at $E=0$ for disorder $W=3$ ($+$), $W=4$
  ($*$), and $W=5$ ($\times$) and $M=201$.  The data are averaged over
  100 samples. The thick (thin) lines indicate data for $U>0$
  ($U<0$). }
\end{figure}

\clearpage
\begin{figure}[th]
\centerline{\epsfig{figure=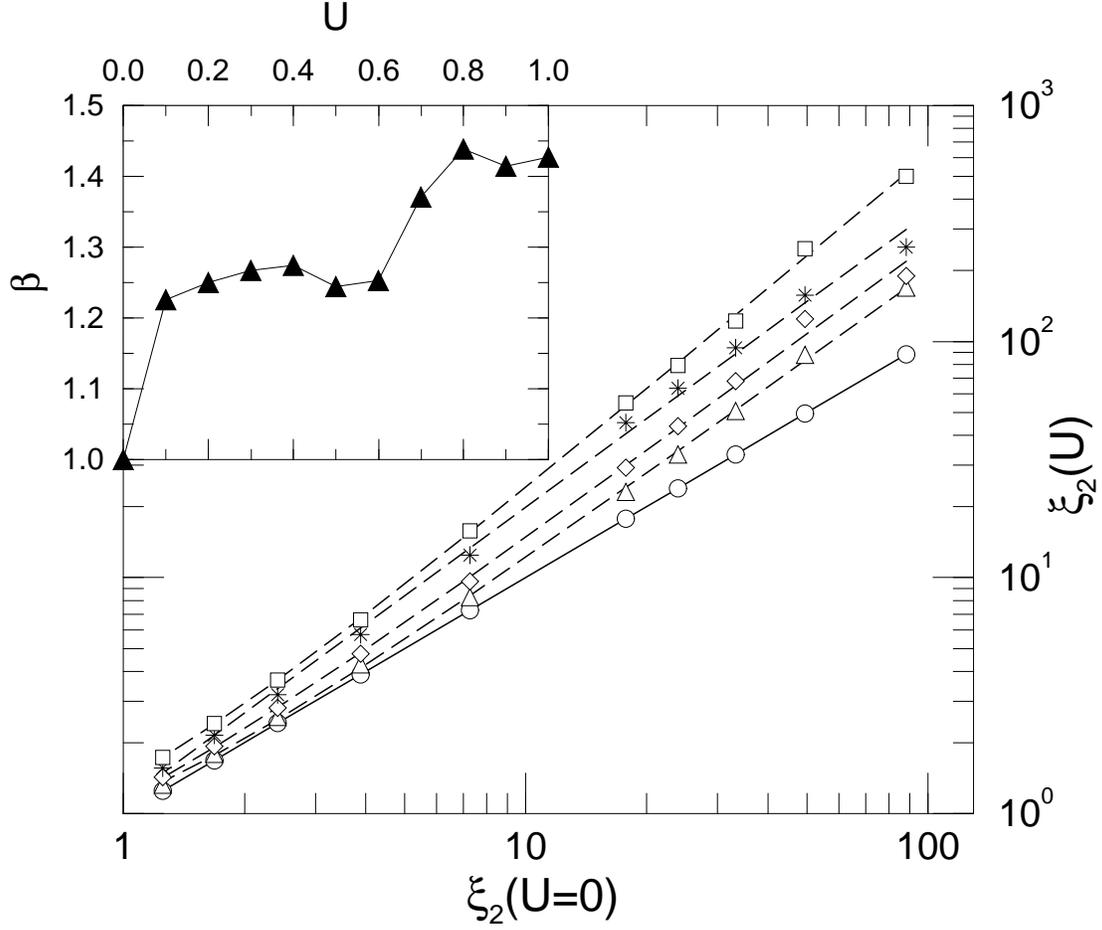,width=\figwidth} }
\caption{\label{fig-iehdm-ps} 
%
  IEH localisation lengths $\xi_2(U)$ after FSS for $U=0$
  ($\bigcirc$), $U=0.1$ ($\triangle$), $U=0.2$ ($\Diamond$), $U=0.5$
  ($*$) and $U=1$ ($\Box$) plotted versus $\xi_2(0)$. The lines are
  fits as in Fig.\ \protect\ref{fig-tipdm-ps}.  Inset: Exponent
  $\beta$ obtained by fitting Eq.\ (\protect\ref{eq-ps}) to the data
  for $U= 0, 0.1, \ldots, 1$.}
\end{figure}


\clearpage
\begin{figure}[th]
\centerline{\epsfig{figure=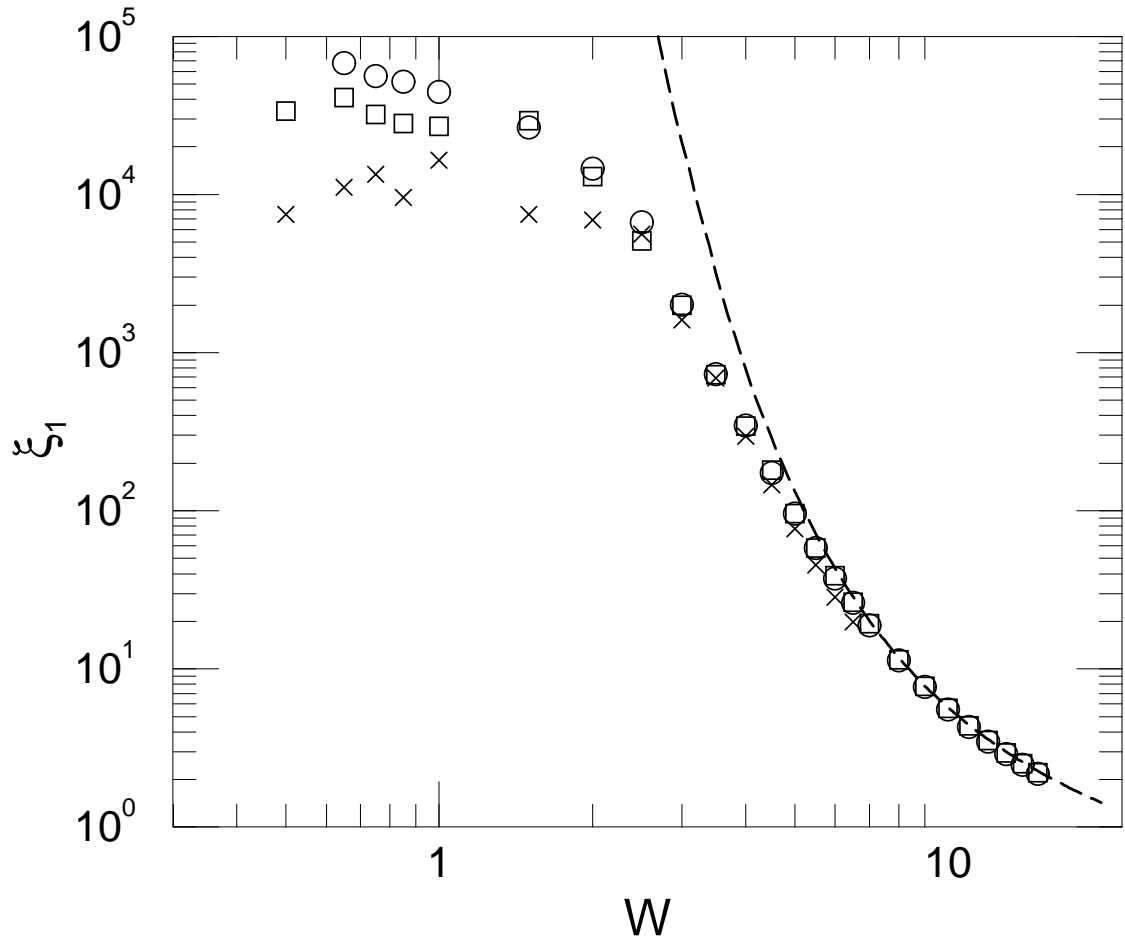,width=\figwidth} }
\caption{\label{fig-a2ddm-xi_u}
%
  SP localisation lengths as in Fig.\ \protect\ref{fig-a2ddm-xi} for
  $U=0$ (-- --, $\bigcirc$) and $\protect\sqrt{2}\ \xi_1$ obtained by
  DM for $U=1$ with additional random ($\Box$) or constant ($\times$)
  potential energies along the diagonal.  }
\end{figure}


\begin{references}
\bibitem{remark} The $\lambda_2$ values obtained here are defined as
  half the $\lambda_2$ values of Refs.
  \cite{FMPW95,RS97,RS98,HZKM98}.
\end{references}
\end{document}